\documentclass[aps,prd,twocolumn,nofootinbib,amsmath,amssymb]{revtex4-1}
\usepackage[dvips]{graphicx}         
\usepackage{mathptmx}                
\usepackage{physics}                 
\usepackage{dcolumn}                 
\usepackage{multirow}                
\usepackage{xcolor}
\usepackage{hyperref}
\hypersetup{,breaklinks=true,colorlinks=true,
linkcolor=blue,citecolor=blue,filecolor=magenta,urlcolor=black}

\long\def\OFF#1{}
\long\def\hj#1{\color{red}#1\color{black}}

\def\be{\begin{equation}} \def\ee{\end{equation}}
\def\bal#1\eal{\begin{align}#1\end{align}}

\def\eps{\varepsilon}
\def\la{\lambda}

\def\om{\omega}
\def\ms{\,M_\odot}
\def\km{\,\text{km}}
\def\mmax{M_\text{max}}
\def\fm3{\,\text{fm}^{-3}}
\def\mfm{\,\text{MeV}\,\text{fm}^{-3}}

\begin{document}

\title{Radial oscillations of proto-neutron stars}


\author{T. T. Sun$^{1}$}
\author{H. Chen$^{1}$}\email{Email:huanchen@cug.edu.cn}
\author{J. B. Wei$^{1}$}
\author{\hbox{Z. Y. Zheng$^{2}$}}

 \author{G. F. Burgio$^{3}$}
 \author{H. -J. Schulze$^{3}$}

\affiliation{
$^{1}$ School of Mathematics and Physics, China University of Geosciences,
Lumo Road 388, 430074 Wuhan, China
\\
$^{2}$ Institute of Astrophysics, Central China Normal University,
Luoyu Road 152, 430079 Wuhan, China
\\
$^{3}$ INFN Sezione di Catania, Dipartimento di Fisica,
Universit\'a di Catania, Via Santa Sofia 64, 95123 Catania, Italy
}

\begin{abstract}
We investigate radial oscillations of proto-neutron stars,
employing equations of state described by the Brueckner-Hartree-Fock theory
or the relativistic mean field model,
and assuming isentropy and fixed lepton fractions for the internal structure.
We calculate the eigenfrequencies and corresponding oscillation functions,
which show different characteristics in different mass regions.
In the low-mass region around $1.4\ms$,
the radial oscillation frequencies are lowered
by large entropy and neutrino trapping,
along with a reduction of the average adiabatic index.
In the region close to the maximum mass,
the fundamental oscillation frequency drops rapidly
and vanishes at the maximum mass,
in accordance with the critical stability criterion
$\partial M /\partial \rho_c = 0$,
as for cold neutron stars.
\end{abstract}



\maketitle

\section{Introduction}

It is well known that neutron stars (NSs) are the most compact objects
in the universe except black holes.
They provide natural laboratories for studying properties of dense matter
with several times the saturation density $\rho_0\simeq\,0.16\fm3$.
The temperature may also be very important in the early stage of a NS,
e.g., in a proto-neutron star (PNS) after a supernova explosion
or in the remnant of a binary neutron star merger (BNSM) event.
PNSs are mainly formed in a Type II supernova explosion \cite{Prakash97}
of a massive star ($M \gtrsim 8\ms$).
PNSs possess extremely dense and hot interiors,
with temperatures exceeding tens of MeV and densities surpassing
the nuclear saturation density by several times \cite{Lu19}.
During the evolution of PNSs,
a large variety of electromagnetic signals and neutrino signals
\cite{Keil95,Pons99}
as well as gravitational signals
\cite{Burgio11,Zhu18,Mu22}
could be emitted and detected.
Through the study of PNSs,
one can thus acquire a deeper understanding of hot/dense nuclear matter (NM)
and NSs.

The first observation of the BNSM event GW170817
has opened a new era of research into NSs \cite{Abbott17}.
The remnant of a BNSM may be a stable NS
\cite{Stergioulas11,Bauswein16,Baiotti17,Janka22}
or a hypermassive NS existing for a short duration
before collapse into a black hole \cite{Baiotti17,Gill19}.
Such a remnant may have a similar dense and hot environment as a PNS
\cite{Dessart08,Ciolfi17,Sumiyoshi21b}.
Finite-temperature effects also affect the stability \cite{Figura21}.
Therefore, the study of PNSs can also shed light
on the remnant of a BNSM.

One key element for the study of PNSs is the equation of state (EOS),
which is still affected by uncertainties of the theory
under the extreme interior conditions of PNSs.
Laboratory and astrophysical observations are thus required to constrain the EOS.
For the static and spherical case,
one can obtain the equilibrium structure of (P)NSs by solving the
Tolman-Oppenheimer-Volkoff (TOV) equations combined with the EOS,
thus predicting mass-radius-central density relations.
So the mass and radius observations impose the most direct constraints
on the EOS.

The most massive cold NSs observed so far are
PSR J0952-0607 with $M=2.35\pm0.17\ms$ \cite{Romani22}
and PSR J0740+6620 with $M=2.08\pm0.07\,\ms$ \cite{Fonseca21}
and measured radius
$13.7^{+2.6}_{-1.5}\km$ \cite{Riley21} or
$12.4^{+1.3}_{-1.0}\km$ \cite{Miller21}.
The analysis of the GW170817 gravitational wave (GW) event
limits the maximum mass of cold NSs to about $2.3\,\ms$
\cite{Shibata17,Margalit17,Rezzolla18,Shibata19,Shao20,Nathanail21}.
These are important constraints on the EOS of cold dense nuclear matter.
Similarly, if the mass and radius of a PNS could be observed,
this would constrain the hot dense nuclear-matter EOS in PNSs
\cite{Zhu18,Lu19,Logoteta22}.

Radial oscillations are the most fundamental oscillation modes,
and can be used to probe the internal structure and stability of stars
\cite{Chandrasekhar64,Chanmugam77,Gondek97}.
The latter determines the collapse to a black hole
\cite{Brillante14,Panotopoulos17,Sotani21,Koliogiannis21,Zhang24}.
Our previous work \cite{Sun21} and other recent works
\cite{Brillante14,Panotopoulos17,Pereira18,Goncalves22,Rau23,Zhang24}
examined the difference of the radial oscillations
between cold hybrid stars and pure NSs.
Refs.~\cite{Pereira18,Goncalves22,Rau23}
investigated the influence on radial oscillations
of a slow conversion rate
during the nuclear matter - quark matter phase transition,
and found it can modify the simple stability condition
$\partial M / \partial \eps_c \geq 0$.

Although radial oscillations cannot be observed directly,
they can couple with non-radial oscillations
that can directly produce GW signals \cite{Soultanis21},
and therefore could be indirectly observed.
Radial oscillations can also modulate a short gamma ray burst (SGRB)
from the hypermassive NS formed after a BNSM,
and the frequency could thus be observed in SGRBs \cite{Chirenti19}.
So far, the radial oscillations of
cold neutron stars, hybrid stars, and quark stars have been widely studied
\cite{Sagun20,Dima21,Lihongbo22},
but the research on radial oscillations of PNSs is limited
\cite{Gondek97,Zhu18,Sotani21},
and we dedicate this work to it.

Immediately after the core bounce of a massive star,
the core shrinks to less than $20\km$, forming a PNS,
which then starts the Kelvin-Helmholtz evolution
in the following several tens of seconds.
During this epoch,
neutrino trapping occurs because the neutrino mean free path
is much smaller than the star's radius \cite{Pons99}.
Neutrinos thus contribute to the pressure and energy density in the star,
and influence the structure and evolution of PNSs
\cite{Prakash97,Logoteta22}.
The Kelvin-Helmholtz epoch usually consists of two major evolutionary stages
\cite{Burrows86,Prakash97,Pons99,Camelio17}:
the deleptonization stage and the cooling stage.
At the beginning of the deleptonization stage,
the PNS has a high-entropy mantle while the core is cooler than the mantle.
Then the excess trapped electron neutrinos diffuse out of the central region,
heating the core,
while the net lepton and proton fractions decrease.
When the entropy per baryon in the center of the PNS reaches its maximum,
about $2k_B$,
the deleptonization stage is finished.
An overall cooling stage follows,
with the entropy as well as lepton fractions in the star steadily decreasing.
According to simulations of the supernova collapse dynamics,
the entropy per baryon in the PNS is approximately constant (isentropic)
during the cooling stage \cite{Thompson02,Burrows12}.
In this work,
we thus use isentropic conditions for the EOS during the cooling stage,
and compare neutrino trapping with constant lepton fractions
$Y_{L_e}=0.4,\ Y_{L_\mu}=0$
with neutrino-free conditions.

Here we further study the influence of temperature and neutrino trapping
on the radial oscillations of PNSs,
based on our previous works on cold NSs
\cite{Sun21,Zheng23}.
There are various theoretical models for the PNS EOS
\cite{Burgio18,Burgio21,Raduta21,Wei21}.
Commonly used models include the liquid-drop-type model
\cite{Lattimer91},
relativistic-mean-field (RMF) models
\cite{Chabanat98,ShenG11,Dutra14,Constantinou14,Shen19,Shen20,Mu22},
Brueckner-Hartree-Fock (BHF) theory
\cite{Burgio10,Lu19,Wei21,Liu22},
and extended nuclear statistical equilibrium models
\cite{Ropke09,Hempel10,Raduta18}.
For an overview, see Ref.~\cite{Burgio21}.
We will use the recent version of a BHF EOS
at finite temperature \cite{Lu19}, 
based on realistic nucleon-nucleon ($NN$) and three-nucleon forces,
and compare with a modern RMF EOS \cite{Shen20}
with a small symmetry energy slope.

The article is organized as follows.
In Sec.~\ref{s:eos} we briefly describe our adopted EOSs in PNSs, i.e.,
the BHF EOS and the Shen RMF EOS.
In Sec.~\ref{s:osc} we introduce the TOV and the Sturm-Liouville
eigenvalue equations for the internal structure and radial oscillations of PNSs.
Numerical results are presented in Sec.~\ref{s:res},
and conclusions in Sec.~\ref{s:end}.
We use natural units $c=\hbar=k_B=G=1$ throughout.

\section{Equation of state for proto-neutron stars}
\label{s:eos}

In this work,
we use either a BHF EOS \cite{Lu19}
or the Shen 2020 RMF EOS \cite{Shen20} for the core of a PNS.
BHF theory provides a non-relativistic microscopic EOS
based on realistic nuclear forces.
The key element to describe the dense nuclear matter
is the in-medium $NN$-interaction $K$-matrix,
which satisfies the well-known Bethe-Goldstone equation \cite{Baldo99}.
In the case of zero temperature,
the probability of nucleons occupying a single-particle state
is described by a step function,
and at finite temperature by the Fermi-Dirac distribution.
Following the Bloch-De Dominicis approach \cite{Bloch1,Bloch2},
for a given density and temperature,
the Bethe-Goldstone equation can be solved self-consistently,
and the free energy density $f$ calculated exactly \cite{Liu22}
or in the simplified ``frozen-correlations" approximation \cite{Burgio10,Lu19}.
In this manuscript we choose the latter approach.

From the free energy density as function of temperature
and particle number densities $\rho_i$,
one can in turn determine various thermodynamic quantities
such as chemical potentials $\mu_i$, pressure $p$, energy density $\eps$,
and entropy density $s$,
\bal
 & \mu_i = {\partial f \over \partial \rho_i} \:,
\\
 & p = \rho^2 {\partial (f/\rho) \over \partial \rho}
     = \sum_i \mu_i \rho_i - f \:,
 \\
 & s = - {\partial f \over \partial T}\:,
  \\
 & \eps = f + Ts\:,
\eal
where the subscript $i=n,p$ and $\rho=\rho_n+\rho_p$.

In this work we use the parameterized free energy density of nuclear matter
obtained in \cite{Lu19}
with the Argonne $V_{18}$ (V18) $NN$ potential \cite{Wiringa95},
supplemented with compatible microscopic three-body forces
\cite{Zuo02,Zhou04,Li08}.
The structure of corresponding PNSs is also discussed in \cite{Lu19}.

The non-relativistic BHF theory provides a quite stiff EOS
with a large sound speed,
even superluminal at very large density.
Therefore we consider also an RMF EOS,
with a sound speed fulfilling the causality requirement.
RMF theory describes nuclear matter based on quantum field theory,
which treats nucleons as quanta of fields and describes their interaction
with exchanged mesons such as $\sigma, \omega, \rho$.
From the corresponding Lagrangian,
the Euler-Lagrange equations for the fields can be derived.
Under the mean-field approximation,
the meson fields are treated as classical fields
and the field operators are replaced by their expectation values.
The energy density, pressure, and other thermodynamic quantities
of nuclear matter can be obtained by solving these equations.
A parametrization, i.e.,
a set of parameters in the Lagrangian is needed to fit experimental data.

In the following, we adopt the Shen 2020 EOS \cite{Shen20},
which is often employed for PNSs and supernova simulations
\cite{Shen02,Shen11,Shen19}.
It adopts the TM1e parametrization \cite{Bao14},
which modifies the density dependence of the symmetry energy
by introducing additional $\omega-\rho$ coupling terms.
Compared to the original TM1 parametrization \cite{Shen98,Shen98b},
TM1e gives smaller symmetry energies,
and is more consistent with current observations of NSs.
In this work we use the EOS table published in CompOSE \cite{compose}.

As stated above,
we consider the PNS stage which is approximately isentropic.
The star contains nucleons and leptons
including neutrons, protons, photons, electrons, muons,
and corresponding neutrinos.
Considering neutrino trapping,
we constrain the hot dense nuclear matter in beta equilibrium,
charge neutral, and combine with fixed lepton numbers,
\bal
& \mu_p + \mu_e = \mu_n + \mu_{\nu_e} \:,
\\
& \mu_p + \mu_\mu = \mu_n + \mu_{\nu_\mu} \:,
\\
& \rho_p - \rho_e - \rho_{\mu} =0\:,
\\
& Y_{L_e} = (\rho_e + \rho_{\nu_e})/\rho \:,
\\
& Y_{L_\mu} = (\rho_\mu + \rho_{\nu_\mu})/\rho \:,
\eal
where $\rho_i$ and $\rho_{\nu_i}$ are the net electron/muon
and corresponding neutrino number densities, respectively.
$Y_{L_i}$ are the lepton fractions,
$Y_{L_e}=0.4$ and $Y_{L_\mu}=0$ for trapped matter.
In absence of neutrino trapping,
$\mu_{\nu_e}=\mu_{\nu_\mu}=0$,
and the lepton fractions $Y_{L_i}$ are not constrained.

We notice that the BHF theory provides only the EOS in the core.
However, the RMF EOS also describes non-homogeneous nuclear matter
in the crust of PNSs, using the Thomas-Fermi approximation.
It assumes that below a certain density threshold,
heavy nuclei can co-exist with a free nucleon gas,
and reduce the free energy of the system.
By minimizing the free energy,
the most stable state of nuclear matter is determined.
Therefore, we adopt the Shen EOS \cite{Shen20} for the low-density crust,
irrespective of the core.
As in similar works \cite{Zhu18,Sotani21},
we assume a PNS surface energy density of $10^5\,\text{g/cm}^3$,
and disregard the surrounding atmosphere.

\section{Hydrostatic equilibrium structure and radial oscillations}
\label{s:osc}

The radial oscillations of PNSs are regarded as adiabatic perturbations
in hydrostatic equilibrium.
The static spherically-symmetric stars are
described by the Schwarzschild metric \cite{Chandrasekhar64}
\be\label{e:ds2}
 ds^2 = e^{\nu(r)}dt^2 - e^{\lambda(r)}dr^2 -
 r^2(d\theta^2 + \sin^2\theta d\varphi^2) \:,
\ee
where $e^{\nu(r)}$ and $e^{\lambda(r)}$ are metric functions.
The static equilibrium configurations of PNSs
are described by the TOV equations \cite{Oppenheimer39,Tolman39}
for pressure $p$ and enclosed gravitational mass $m$
and baryonic mass $m_B$,
\bal\label{dpdr}
 p' &= -\frac{e^\la}{r^2} (p+\eps)(m+4\pi r^3p) \:,
\\
 m' &= 4\pi r^2\eps \:,
\\
 {m_B}' &= 4\pi r^2 e^\la \rho m_n \:,
\eal
where $p' \equiv dp/dr$ etc.\ and $m_n$ is the neutron mass.
The corresponding metric functions are
\bal
 e^{\la(r)} &= \frac{1}{1-2m/r} \:,
\\
 \nu(r) &=
 -2 \int_r^\infty\!\!d\tilde r\,
 \frac{e^{\la(\tilde{r})}}{\tilde r^2} \left( m + 4\pi \tilde r^3 p \right) \:.
\eal
By combining the TOV equations with the EOS $p(\eps)$
one can compute the mass-radius relation of PNSs and
the various distributions of thermodynamic quantities in PNSs.
The equations are solved with the central conditions
$m(r=0)=0$ and $p(r=0)=p_c$,
integrating until the surface $p(R)=0$.
Correspondingly the PNS gravitational mass is $M=m(R)$.

In the case of small radial perturbations of PNSs,
the metric in Eq.~(\ref{e:ds2}) would be time dependent.
Following \cite{Chanmugam77,Chanmugam92,Sagun20}
we introduce the fluid perturbations
$\xi\equiv\Delta r/r$,
where $\Delta r$ is the radial displacement,
and the Lagrangian perturbation of the pressure
$\eta\equiv\Delta p/p$.
By separation of the time-dependent factor $e^{i\om t}$ of $\xi$ and $\eta$,
we obtain the equations for the $r$-dependent oscillation amplitudes
\cite{Chanmugam77,Gondek97},
\bal
 \xi' = &
 - \frac{p'\xi}{p+\eps}
 - \frac{1}{r} \Big( 3\xi + \frac{\eta}{\Gamma} \Big) \:,
\label{e:xi}
\\
 \eta' = &\hskip1mm
 \frac{\xi}{p} \bigg[ \om^2 r e^{\lambda-\nu} (p+\eps) - 4p'
 + \frac{r p'^2}{p+\eps}
 - 8\pi r e^\lambda(p+\eps)p \bigg]
\nonumber\\
 & - \eta \bigg[ \frac{p'\eps}{p(p+\eps)}
                 + 4\pi r e^\lambda(p+\eps) \bigg] \:,
\label{e:dp}
\eal
where
$\om=2\pi f$ is the angular frequency of radial oscillation
with the eigenfrequency $f$,
and
\be\label{e:gamma}
 \Gamma =
 \frac{\partial \ln{p}}{\partial \ln\rho}\Bigg|_{s} \:
 = \left(1+\frac{\eps}{p}\right) v_s^2
\ee
is the adiabatic index 
with the squared speed of sound
\be\label{e:vs}
 v_s^2 = (\partial p/ \partial \eps)_{s,\text{equil.}} \:
\ee
for the isentropic EOS in beta equilibrium.

It is believed that the time scale of weak interaction
is much larger than that of the radial oscillations, 
and thus the components are frozen during oscillations \cite{Pereira18},
$v_s^2 = (\partial p/ \partial \eps)_{s,\text{frozen}}$.
We compared the results by using the two procedures
and found that there is little difference,
confirming the results of \cite{Gondek97}.

The solution of the radial oscillation equations
requires two boundary conditions.
One is the condition of regularity at \hbox{$r=0$}
\cite{Chanmugam77,Gondek97},
\be\label{e:r0}
 [\eta + 3\Gamma\xi](r=0) = 0 \:,
\ee
and the other is the vanishing of the perturbation of the pressure
at the surface,
\be\label{e:rR}
 0 = \Delta p(r=R) = - \xi p \left[ e^{\lambda(R)}
 \left( \frac{\omega^2 R^3}{M} + \frac{M}{R}\right) + 4 \right] \:.
\ee

Eqs.~(\ref{e:xi},\ref{e:dp})
are the Sturm-Liouville eigenvalue equations for $\om$.
The solutions provide the discrete eigenvalues $\om_i^2$.
For a given PNS, they can be ordered as
$\om_1^2 < \om_2^2 < \ldots < \om_n^2$,
where $n-1$ is the number of nodes of $\xi(r)$ and $\eta(r)$.
Negative $\om^2$ indicates an unstable perturbation
and thus $\om_1^2=0$ is the critical condition for the stability of PNSs
under radial perturbations.

\section{Numerical results and discussion}
\label{s:res}

After constructing the isentropic EOS with neutrino trapping in beta-equilibrium,
we solve numerically the TOV and radial oscillation equations for PNSs,
obtaining their properties of structure and corresponding radial oscillations.
In the following, we show and discuss the numerical results,
comparing with those for cold NSs and hot NSs without neutrino trapping.

\begin{figure}[t]
\vskip-2mm
\centerline{\includegraphics[width=0.5\textwidth]{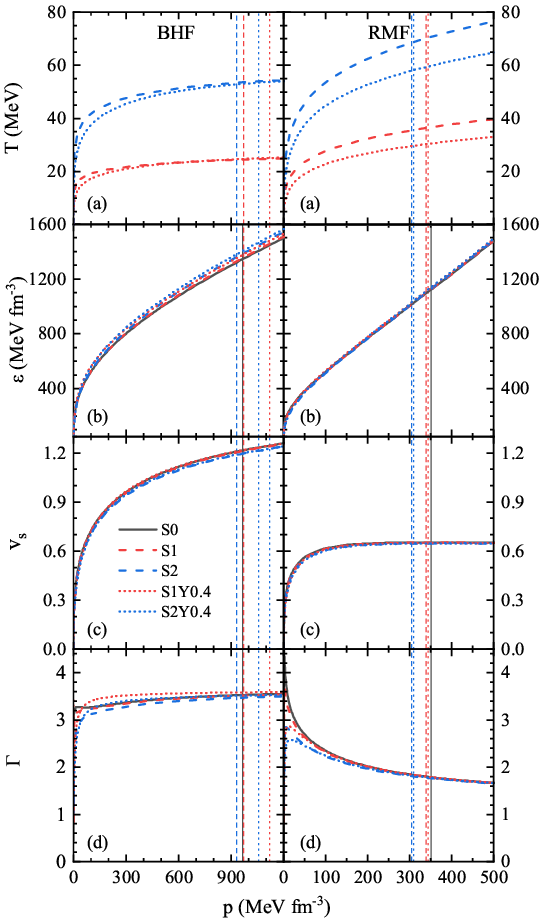}}
\vskip-2mm
\caption{
The temperature (a),
energy density (b),
sound speed (c), and
adiabatic index (d)
of (P)NSs as functions of pressure
with BHF (left panels) and RMF (right panels) EOS,
for various values of $S/A$ and lepton fractions,
see the text for a detailed description of the notation.
The vertical lines indicate $\mmax$ configurations.
Note the different $p$ axes.
\hj{}
}
\label{f:eos}
\end{figure}

\begin{figure}[t]
\centerline{\includegraphics[width=0.51\textwidth]{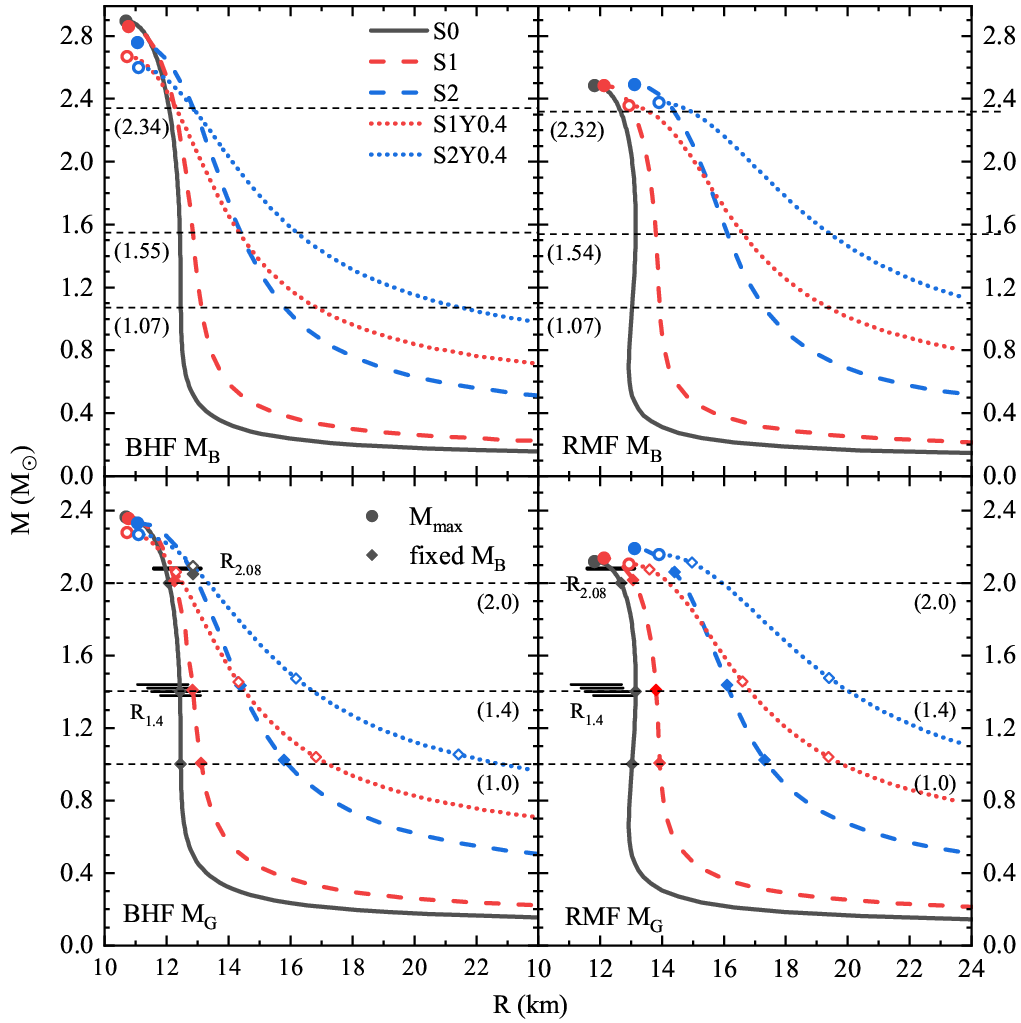}}
\vskip-3mm
\caption{
The baryonic and gravitational mass -- radius relations of (P)NSs
obtained with BHF (left panels) and RMF (right panels) EOSs.
The notation and the $\mmax$ markers correspond to Fig.~\ref{f:eos}.
Configurations of $M_G$ for fixed values of $M_B$ 
are indicated in the lower panels.
The horizontal black bars indicate the limits on
$R_{2.08}$ and $R_{1.4}$
obtained in the combined NICER+GW170817 data analyses
of \cite{Miller21,Pang21,Raaijmakers21}.
}
\label{f:mr}
\end{figure}

\subsection{Equation of state}

For our isentropic EOS,
we first present in Fig.~\ref{f:eos} the
temperature (a),
energy density (b),
sound velocity (c), and
adiabatic index (d),
as functions of pressure.
The results in the left (right) panels are obtained with BHF (RMF) EOS.
Results for $S/A=0,1,2$
(denoted as ``S0, S1, S2")
and $Y_{L_e}=0.4$, $Y_{L_\mu}=0$ for neutrino trapping
(denoted as ``Y0.4")
are compared.
The static $\mmax$ configurations are indicated for all cases by vertical lines.
It can be seen that for the BHF EOS,
finite temperature and trapping lead to a substantial increase of the
maximum central pressure,
whereas for the RMF EOS the effect is mostly a reduction.
This and the related change of the maximum masses
have been analyzed at length in \cite{Wei21,Liu22}.

The upper panels show that temperatures of several tens MeV
can be reached in dense matter,
slightly higher for the RMF EOS.
Neutrino trapping reduces the temperature,
due to the presence of more species.
As the related thermal energies are small,
the entropy has a visible impact (reduction) on the EOSs only at low pressure.
Neutrino trapping further enhances this effect,
because more Fermi seas are available.

Due to the different curvatures of the EOS $\eps(p)$,
the sound speed in panels (c) is quite different for both EOSs,
increasing to large values with the BHF and reaching a plateau
$v_s\approx0.6$ for the RMF.
This has consequences for the adiabatic index in panels (d),
which remains large for the BHF but decreases with pressure for the RMF model.
The qualitative impacts of entropy and neutrino trapping
on the adiabatic index are similar,
significant suppression at low pressure,
as was also found in \cite{Casali10,Lu19,Figura20,Stone21}.

\subsection{Stellar structure}

Then in Fig.~\ref{f:mr} we show the mass -- radius relations of PNSs,
with the same legend as in Fig.~\ref{f:eos}.
The lower panels show the gravitational mass as function of radius,
while the upper panels show the corresponding baryonic mass,
useful for studying the evolution of PNSs.
Some characteristic static properties of the $\mmax$ configurations
of (P)NSs with various EOSs are also listed in Table~\ref{t:res}.
The maximum gravitational mass of cold NS is $2.36 (2.12)\,\ms$
and the corresponding radius is $10.7 (11.8)\,$km with the BHF (RMF) EOS,
respectively.
Furthermore, black horizontal bars
indicate recent limits on the radii $R_{2.08}$ and $R_{1.4}$,
imposed by a combined (strongly model-dependent) analysis
of NICER and GW170817 observations
\cite{Abbott17,Abbott18},
namely
$R_{2.08}=12.35\pm0.75\km$ \cite{Miller21},
and in particular
$R_{1.4}=12.45\pm0.65\km$ \cite{Miller21}, 
$11.94^{+0.76}_{-0.87}\km$ \cite{Pang21}, and
$12.33^{+0.76}_{-0.81}\km$ or
$12.18^{+0.56}_{-0.79}\km$ \cite{Raaijmakers21}.
The BHF EOS is well compatible with these constraints
\cite{Wei20,Burgio21,Sun21},
and also its maximum mass $\mmax \approx 2.36\ms$
exceeds the currently known lower limits,
while the RMF EOS is only marginally compatible with the $R_{1.4}$ range,
and also with the recent measurement of $M=2.35\pm0.17\ms$
for PSR J0952-0607 \cite{Romani22}.

\OFF{
fulfill the constraints from present observations on NS mass and radius,
in particular the recent mass-radius results of the NICER mission
for the pulsars J0030+0451 \cite{Riley19,Miller19}
and J0740+6620 \cite{Riley21,Miller21,Pang21,Raaijmakers21}.
Some theoretical analyses of the GW170817 event indicate also an upper limit
on the maximum mass of $\sim$2.2--2.4$\,\ms$
\cite{Shibata17,Margalit17,Ruiz18prd,Rezzolla18,Shibata19},
with which the V18 EOS would be compatible as well.
However, those are very model dependent,
in particular the still to-be-determined temperature dependence of the EOS
\cite{Khadkikar21,Bauswein21,Figura21,Liu22}.
}

Regarding hot stars,
with the BHF EOS
the maximum gravitational mass of PNSs is slightly smaller than that of cold NSs.
For S2, the maximum mass decreases by about $0.03\,\ms$
and the corresponding radius increases by about $0.4\,$km.
For the RMF EOS instead, $\mmax$ increases by about $0.08\,\ms$
and the radius by about $1.3\,$km.
For an extended discussion, see \cite{Wei21,Liu22}.
Including neutrino trapping, however,
the maximum gravitational decreases by a few percent in all cases.

Compared to cold NSs,
the maximum baryonic mass of PNSs also decreases in all configurations,
mainly due to neutrino trapping.
During the evolution of a PNS,
its baryonic mass is usually assumed to be constant.
Therefore, a PNS with a large mass will always be stable
during evolution to a cold NS, without a delayed collapse.
We notice that PNSs with fixed baryonic mass
have also almost the same gravitational mass.
For example, for a cold NS with $M_G=1.4\,\ms$
we obtain $M_B=1.55~(1.54)\,\ms$ with BHF (RMF) EOS,
as shown in the figure.
The corresponding gravitational masses of (P)NSs are shown in the lower panel
as diamond markers.
One can see that they are all very close to $1.4\,\ms$.

As shown in Fig.~\ref{f:eos},
the increase of entropy or neutrino trapping
has a relatively large effect on the EOS at low pressure,
and correspondingly causes a remarkable increase of the radius
for a low-mass PNS, e.g.,
a $1.4\,\ms$ PNS with S2Y0.4
has a radius about 50\% larger than the cold NS with the same mass
for both EOSs.

\begin{table}[t]
\caption{
Properties of (P)NSs with maximum mass.
\hj{}
}
\begin{ruledtabular}
\begin{tabular}{c|c|rrrrr}
 \multicolumn{2}{c|}{Model} & $M_G/\!\ms$ & $M_B/\!\ms$ &
 $R$(km) & $\rho_c$($\fm3$) & $p_c$($\mfm$) \\
\hline
\multirow{5}{*}{BHF}
 & S0     & 2.36 & 2.89 & 10.7 & 0.990 &  967 \\
 & S1     & 2.36 & 2.86 & 10.8 & 0.990 &  971 \\
 & S2     & 2.33 & 2.76 & 11.1 & 0.975 &  931 \\
 & S1Y0.4 & 2.28 & 2.67 & 10.7 & 1.030 & 1120 \\
 & S2Y0.4 & 2.27 & 2.60 & 11.1 & 1.010 & 1057 \\
 \hline
 \multirow{5}{*}{RMF}
 & S0     & 2.12 & 2.48 & 11.8 & 0.910 &  351 \\
 & S1     & 2.14 & 2.48 & 12.1 & 0.884 &  339 \\
 & S2     & 2.19 & 2.49 & 13.1 & 0.807 &  305 \\
 & S1Y0.4 & 2.10 & 2.36 & 12.9 & 0.871 &  344 \\
 & S2Y0.4 & 2.16 & 2.38 & 13.9 & 0.801 &  310 \\
\end{tabular}
\end{ruledtabular}
\label{t:res}
\end{table}

\begin{figure}[t]
\centerline{\includegraphics[width=0.51\textwidth]{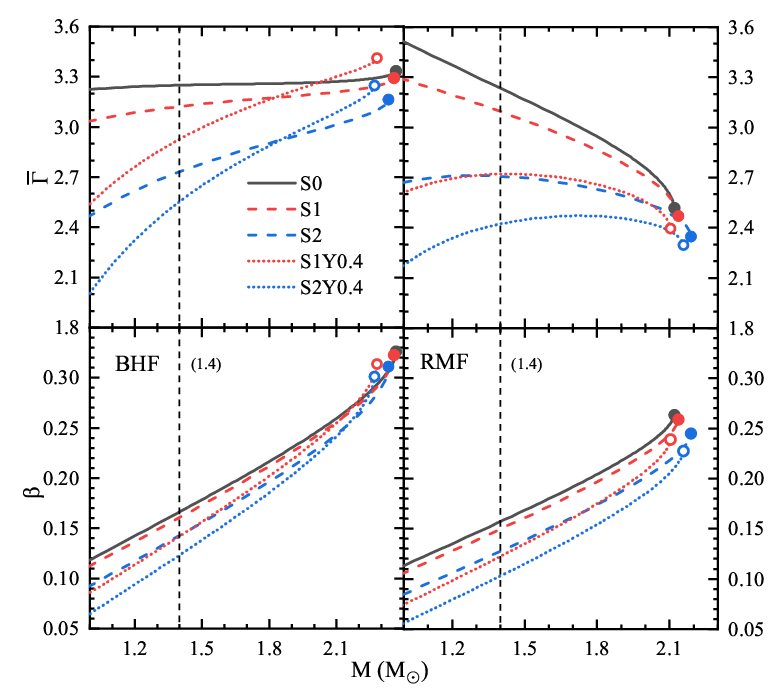}}
\vskip-3mm
\caption{
The average adiabatic index $\bar\Gamma$, Eq.~(\ref{e:ave}),
and compactness parameter $\beta=M/R$
of (P)NSs with various EOSs.
The notation is as in Fig.~\ref{f:eos}.
}
\label{f:gammaave}
\end{figure}

The adiabatic index, Eq.~(\ref{e:gamma}),
is an important factor affecting radial oscillations,
which appears in the oscillation equation (\ref{e:xi})
but not in the TOV equations for NS structure.
Ref.~\cite{Chandrasekhar64} suggested the adiabatic index as a critical quantity
determining the stability of NSs.
The discussion was extended to the average adiabatic index defined as
\be\label{e:ave}
 \bar\Gamma \equiv
 \frac{\int_0^R dr\, r^2 e^{(\lambda+3\nu)/2}p \,\Gamma}
      {\int_0^R dr\, r^2 e^{(\lambda+3\nu)/2}p \phantom{\,\Gamma}} \:,
\ee
and its critical values for stability increase with compactness parameters
$\beta={M}/{R}$ or $p_c/\eps_c$
\cite{Merafina89,Negi01,Moustakidis17}.
We show this quantity in Fig.~\ref{f:gammaave}
together with the compactness parameter $\beta$ of PNSs
as functions of gravitational mass $M$.
The $\beta(M)$ curves simply reflect the behavior of the $M(R)$ relations,
with the BHF EOS predicting more compact stars.
For cold NSs, $\bar\Gamma$ increases (decreases) with increasing mass
for the BHF (RMF) EOS,
and is much smaller close to the maximum mass for the latter.
Correspondingly, the lower maximum compactness with RMF EOS
is qualitatively consistent with the conclusion of \cite{Moustakidis17}
on the relation between the critical adiabatic index and compactness.
In the low-mass region around $M_G=1.4\,\ms$,
the impact of entropy and neutrino trapping are visible,
reducing both the average adiabatic index and compactness parameter,
but the effect becomes rather small close to the maximum mass.

\begin{figure}[t]
\centerline{\includegraphics[width=0.51\textwidth]{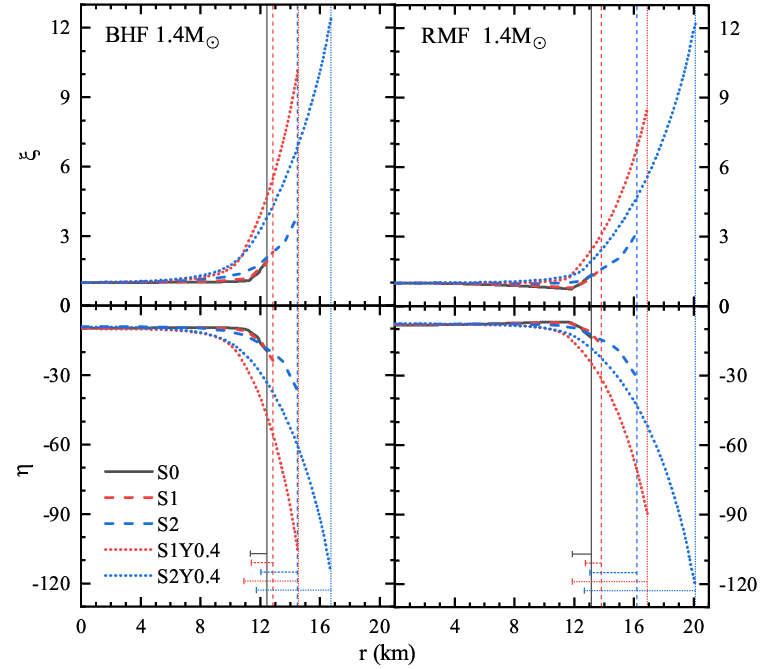}}
\vskip-3mm
\caption{
The eigenfunctions of the fundamental mode for $1.4\ms$ PNSs with various EOSs.
The notation is as in Fig.~\ref{f:eos}.
The vertical lines indicate the radius $R$
and the horizontal bars the extension of the crust.
}
\label{f:internal}
\end{figure}

\subsection{Stellar oscillations}

Fig.~\ref{f:internal} shows the radial oscillation amplitudes $\xi$ and $\eta$
of the fundamental mode in a $1.4\,\ms$ (P)NS
with the conventional normalization $\xi(r=0)=1$.
Due to the difference of $\bar\Gamma$ between BHF and RMF EOS,
the values of $\eta(0)$ are slightly larger in magnitude for the former.
The amplitudes have opposite sign,
as a positive $\Delta r>0$ implies a pressure reduction $\Delta p<0$
and vice versa.
As expected, there is no node in the amplitudes.
All amplitudes increase strongly towards the surface,
facilitated by the more dilute matter.
Therefore,
oscillations occur mainly in the outer part of the star
(the range of the crust is indicated)
and are driven by the `low-density' EOS,
especially for the trapped configurations.

\begin{figure}[t]
\centerline{\includegraphics[width=0.51\textwidth]{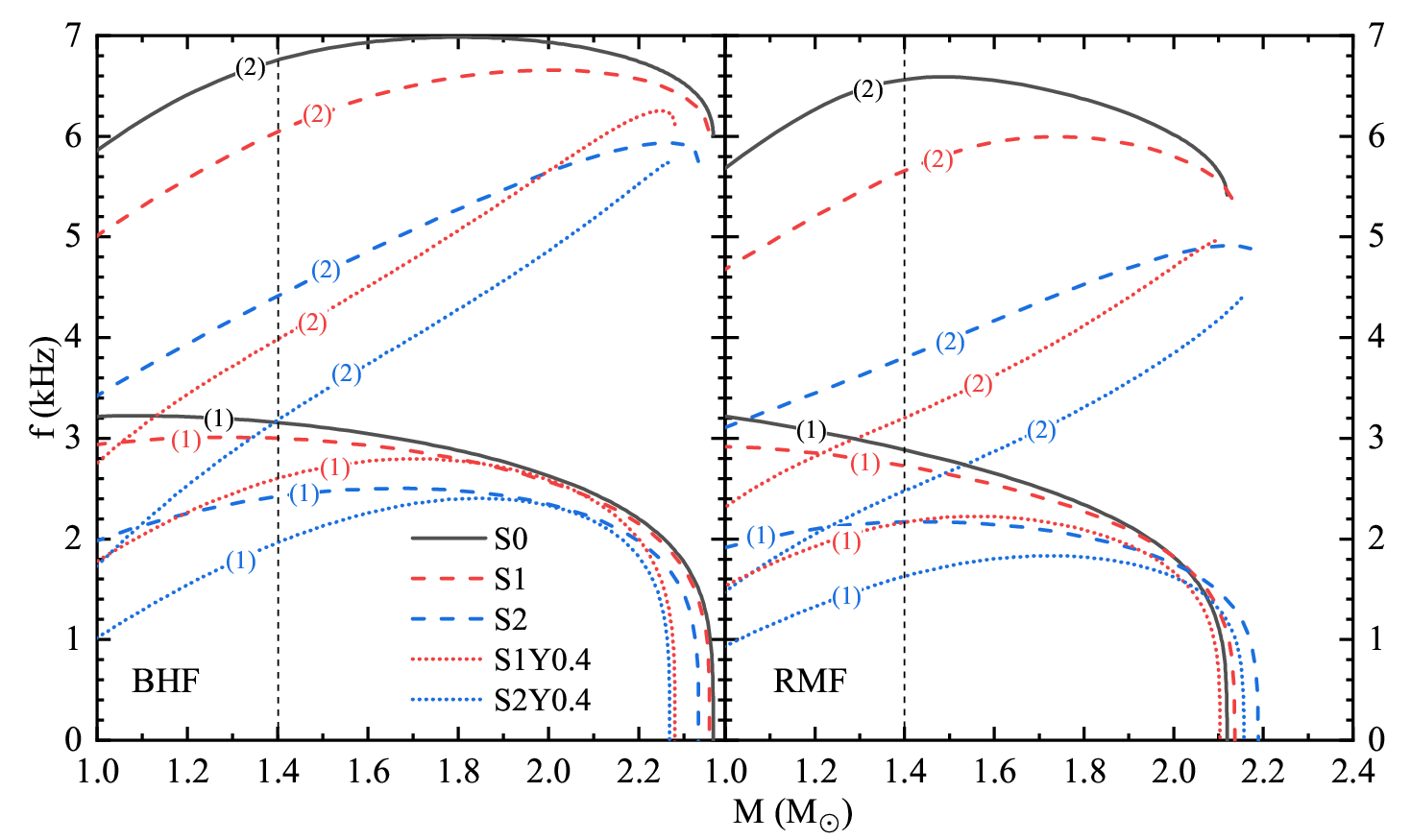}}
\vskip-4mm
\caption{
The radial oscillation frequencies of the first two modes of PNSs
as a function of gravitational mass.
The notation is as in Fig.~\ref{f:eos}.
}
\label{f:fm}
\end{figure}

Fig.~\ref{f:fm} shows the radial oscillation frequencies $f_{1,2}$
of the first two modes as functions of the (P)NS mass.
As stated above,
the radial oscillations are affected by the adiabatic index $\bar\Gamma$.
However, the relations appear differently in two cases.
One case is that for large oscillation frequencies,
such as $f_1$ for moderate masses and $f_2$ in the whole mass region.
For moderate masses the behavior of $f_1$ is very similar
to that of $\bar\Gamma$.
With increasing entropy and trapping $\bar\Gamma$ decreases for both EOSs,
as radii increase and matter becomes more dilute on average.
As a consequence, $f_1$ also decreases visibly.
Quantitatively, for a 1.4$\ms$ object,
$f_1$ of a S2Y0.4 PNS decreases about $1/3$ compared with that of a cold NS.
$f_2$ decreases by even more than 50\%.
Consequently, the quantity $\delta\!f_1 \equiv f_2-f_1$,
which is often discussed in asteroseismology \cite{Sagun20},
is also largely reduced.
Similar suppression can also be found for the higher-mode frequencies,
listed in Table~\ref{t:f}.

The other case is for small oscillation frequencies, i.e.,
$f_1$ in the critical region of stability.
Here the behavior of $f_1$ is quite different,
as it decreases quickly with increasing mass,
and vanishes at the maximum mass.
Therefore, in our model the stability condition
$\partial M / \partial \rho_c > 0$
corresponds indeed to the onset of unstable oscillations $f_1=0$
\cite{Harrison65,Shapiro,Gondek97}.
This is natural since we use the adiabatic index as in
Eqs.~(\ref{e:gamma},\ref{e:vs})
completely determined by the EOS,
which never deviates from the equilibrium configurations during oscillations.
No other information beyond that appearing in the structure equations
is needed for solving the oscillation equations,
which guarantees the consistency between the two sets of equations.
Some exceptions to this criteria are due to
either deviations from the equilibrium configurations during oscillations
\cite{Pereira18,Goncalves22,Rau23,Ghosh24}
or decoupling of the oscillations of different components of NSs
\cite{Arbanil15,Jimenez21,Zhen24}.

\begin{table}[t]
\caption{
The radial oscillation frequencies $f_n\,$[kHz] of $1.4\,\ms$ (P)NSs
with various EOSs.}
\begin{ruledtabular}
\begin{tabular}{c|c|cccccc}
\multicolumn{2}{c|}{$n$} & 1 & 2 & 3 & 4 & 5 & 6 \\
\hline
\multirow{5}{*}{BHF}
 & S0     & 3.15 & 6.76 & 8.60 & 10.08 & 11.42 & 14.04 \\
 & S1     & 3.00 & 6.03 & 7.84 &  9.50 & 10.72 & 12.62 \\
 & S2     & 2.42 & 4.41 & 6.25 &  7.60 &  8.61 &  9.92 \\
 & S1Y0.4 & 2.60 & 3.98 & 5.41 &  6.81 &  8.03 &  9.27 \\
 & S2Y0.4 & 1.97 & 3.18 & 4.27 &  5.35 &  6.39 &  7.37 \\
 \hline
 \multirow{5}{*}{RMF}
 & S0     & 2.90 & 6.55 & 7.87 &  9.07 & 10.85 & 12.95 \\
 & S1     & 2.71 & 5.68 & 7.18 &  8.49 & 10.04 & 11.43 \\
 & S2     & 2.17 & 3.81 & 5.52 &  6.40 &  7.34 &  8.57 \\
 & S1Y0.4 & 2.16 & 3.19 & 4.35 &  5.54 &  6.55 &  7.44 \\
 & S2Y0.4 & 1.62 & 2.46 & 3.29 &  4.15 &  4.97 &  5.70 \\
\end{tabular}
\end{ruledtabular}
\label{t:f}
\end{table}

\begin{figure}[t]
\centerline{\includegraphics[width=0.5\textwidth]{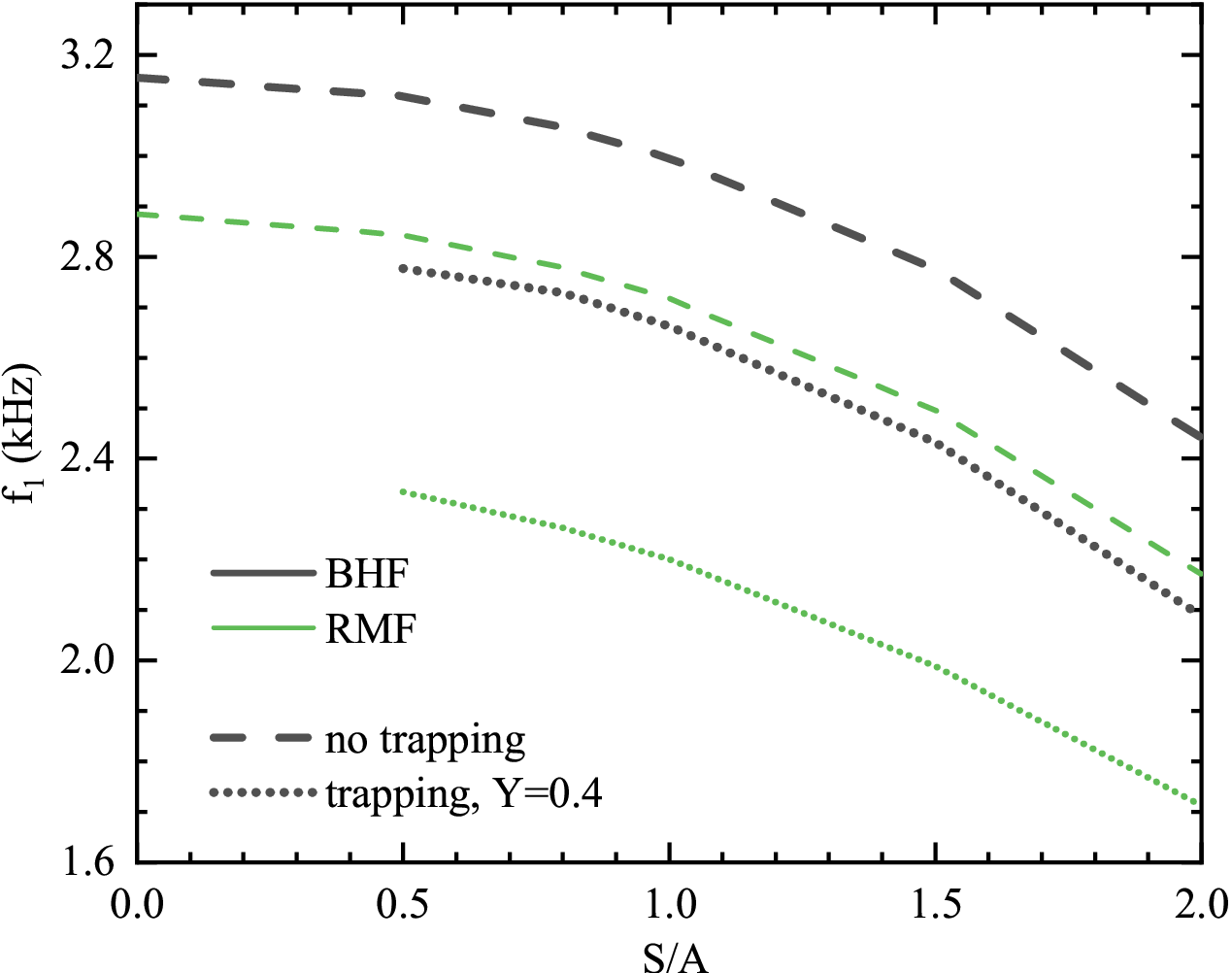}}
\vskip-3mm
\caption{
The fundamental radial oscillation frequency of PNSs
with BHF (thick black lines) or RMF (thin green lines) EOSs
for fixed $M_B$ corresponding to $M_G=1.4\,\ms$.
}
\label{f:fs}
\end{figure}

To investigate the influence of the PNS evolution
on the radial oscillations,
Fig.~\ref{f:fs} shows $f_1$ as a function of entropy per baryon
for a PNS with constant baryonic mass
corresponding to $M_G=1.4\,\ms$,
as shown in Fig.~\ref{f:mr}.
In the early stages of evolution,
the internal entropy of the star is relatively high and trapping is present,
thus the frequency is relatively small.
With decreasing entropy and trapping during the evolution,
$f_1$ increases up to the value of the cold NS.
The figure shows only curves with fixed trapping condition Y0.4,
whereas physically with decreasing temperature
the neutrino free path becomes larger,
neutrinos will escape from the star,
and the frequency increase to the one corresponding to no trapping.
Since the value of $f_1$ depends sensitively on the trapping condition
(see also Table~\ref{t:f}),
its measurement could be a means to determine the actual value of $Y_L$
empirically.

\section{Conclusions}
\label{s:end}

PNSs are high-temperature and lepton-rich objects,
quite different from ordinary low-temperature, neutrino-free NSs.
Consequently, their structure and other properties
are different from cold NSs.
We focused on the radial oscillations and stability of PNSs in this work.
We constructed models of PNSs using the finite-temperature EOS
described by the BHF or RMF theory.
The latter features a softer EOS and lower adiabatic index at large pressure.
Consequently other observables also depend strongly on the EOS.
The internal structure of PNSs is investigated
with a simple environmental approximation:
isentropy and neutrino trapping.

The effect of finite temperature on the maximum gravitational masses of PNSs
depends on the EOS and might be an increase or decrease,
whereas trapping always decreases $\mmax$.
Significant structural differences between PNSs and cold NSs
are present in the small-mass region.
Here the radii of PNSs are much larger than those of cold NSs
due to both thermal and neutrino trapping effects.
This causes a change of the average adiabatic index
and a corresponding significant reduction of radial oscillation frequencies.
For large masses instead,
$f_1$ drops rapidly and vanishes at the maximum mass.
During the PNS evolution,
the oscillation frequency increases significantly,
which might provide a means to obtain information on the state and composition
of matter during this period.

Future work might address several shortcomings of the present one:
We adopted the simple isentropic and constant-lepton-number approximation,
but it is necessary to consider a more realistic environment,
consistent with PNS numerical simulations.
Furthermore, it is possible that different states of matter,
e.g., hyperons or quarks,
may appear in the inner core of PNSs,
which would also significantly influence
their structure and radial oscillations.

\begin{acknowledgments}

We acknowledge financial support from the
National Natural Science Foundation of China
(Grant No. 12205260).

\end{acknowledgments}

\newcommand{\aap}{Astron. Astrophys.}
\newcommand{\apjl}{Astrophys. J. Lett.}
\newcommand{\epja}{Euro. Phys. J. A}
\newcommand{\mnras}{Mon. Not. R. Astron. Soc.}
\newcommand{\npa}{Nucl. Phys. A}
\newcommand{\ptep}{Prog. Theor. Exp. Phys.}
\bibliographystyle{apsrev4-1}
\bibliography{pnsro}

\end{document}